\documentclass[a4paper]{article}

\usepackage{INTERSPEECH2021}

\usepackage{amsmath,graphicx}

\usepackage{subfigure}
\usepackage{amssymb,bm}
\usepackage{textcomp}

\usepackage{epsfig,epstopdf}
\usepackage{stfloats,multirow,booktabs,color,ifpdf,float,array}

\usepackage{amsmath,graphicx}

\usepackage{bm}
\usepackage{amsfonts}
\usepackage{psfrag}
\usepackage{algorithm,algorithmic}
\usepackage{tabu}

\title{Variational Auto-Encoder Based Variability Encoding \\ for Dysarthric Speech Recognition}
\name{Xurong Xie$^{1}$, Rukiye Ruzi$^{1}$, Xunying Liu$^{2}$, Lan Wang$^{1}$}
\address{
  $^1$Shenzhen Institutes of Advanced Technology, Chinese Academy of Sciences, Shenzhen, China\\
  $^2$Chinese University of Hong Kong, Hong Kong, China}
\email{xr.xie@siat.ac.cn, rkym.rouzi@siat.ac.cn, xyliu@se.cuhk.edu.hk, lan.wang@siat.ac.cn}

\begin{document}

\maketitle

\ninept
\begin{abstract}
  Dysarthric speech recognition is a challenging task due to acoustic variability and limited amount of available data. Diverse conditions of dysarthric speakers account for the acoustic variability, which make the variability difficult to be modeled precisely. This paper presents a variational auto-encoder based variability encoder (VAEVE) to explicitly encode such variability for dysarthric speech. The VAEVE makes use of both phoneme information and low-dimensional latent variable to reconstruct the input acoustic features, thereby the latent variable is forced to encode the phoneme-independent variability. Stochastic gradient variational Bayes algorithm is applied to model the distribution for generating variability encodings, which are further used as auxiliary features for DNN acoustic modeling. Experiment results conducted on the UASpeech corpus show that the VAEVE based variability encodings have complementary effect to the learning hidden unit contributions (LHUC) speaker adaptation. The systems using variability encodings consistently outperform the comparable baseline systems without using them, and obtain absolute word error rate (WER) reduction by up to 2.2\% on dysarthric speech with ``Very low'' intelligibility level, and up to 2\% on the ``Mixed'' type of dysarthric speech with diverse or uncertain conditions.
\end{abstract}

\section{Introduction}
\label{sec:intro}
\setlength{\intextsep}{4pt plus 1pt minus 1pt}
\setlength{\textfloatsep}{4pt plus 1pt minus 1pt}
\setlength{\abovecaptionskip}{4pt plus 1pt minus 1pt}
\setlength{\abovedisplayskip}{4pt plus 1pt minus 1pt}
\setlength{\belowdisplayskip}{4pt plus 1pt minus 1pt}

Dysarthria refers to a set of neuromuscular control that impair the physical production of speech. The underlying causes of dysarthria include neurological conditions such as Parkinson disease \cite{scott1983speech}, cerebral palsy \cite{2000Tara}, and brain damages due to stroke or head injuries. The consequence of such disorder includes weakness, slowing, incoordination, altered muscle tone and inaccuracy of oral and vocal movements \cite{Enderby2013}, which result in speech with abnormal characteristics in quality as well as reduced intelligibility. Meanwhile, dysarthria is often associated with physical disability, so speech-driven assistive technology can be beneficial for dysarthric people using speech as an interface for communication \cite{hawley2012voice} or for enabling them to control physical devices \cite{hawley2007speech}. However, commercial automatic speech recognition (ASR) systems trained with normal speech are improper to be directly utilized for dysarthric speech \cite{doyle1997dysarthric,young2010difficulties} due to a large mismatch between training and testing. Sparseness of suitable data is another challenge for ASR system development.

Studies has been carried out for developing dysarthric speech recognition system by constructing new datasets \cite{Kim2008a,rudzicz2012torgo}, or by using data augmentation techniques to deal with data sparsity \cite{jaitly2013vocal,cui2015data,vachhani2018data,genginvestigation}. Variability of dysarthric speech has been modeled by different manners. Speech tempo in signal domain \cite{vachhani2018data} and feature domain \cite{xiong2019phonetic} can be modified to reduce mismatch between dysarthric and normal speech. In \cite{rudzicz2010articulatory,xiong2018deep}, articulatory knowledge is used to reduce inter-speaker variability. Visual information can be used to reduce variability in DNN models \cite{liu2020exploiting}.
Bottleneck features have been extracted using a large amount of data from normal speaker through DNNs \cite{christensen2013combining} or convolutive bottleneck network \cite{nakashika2014dysarthric} to improve dysarthric speech recognition.
Adaptation techniques are also popular for tackling such variation.
Combination of HMM state transition interpolation and maximum a posterior adaptation is exploited to model intra-speaker variability in \cite{sharma2010state}.
Speaker adaptive training \cite{sehgal2015model} is shown to be useful for annihilating inter-speaker variations in dysarthric corpus.
Pronunciation lexicon adaptation \cite{mengistu2011adapting,christensen2013learning} shows effectiveness in a large vocabulary task for dysarthric speakers.
Recently, sequence-to-sequence models including listen, attend and spell (LAS)~\cite{shor2019personalizing,takashima2019end,wang2021improved}, RNN-transducer~\cite{shor2019personalizing}, transformer~\cite{lin2020staged}, and QuartzNet~\cite{wang2021improved} have been used to model dysarthric speech.

Some of the aforementioned studies treat the dysarthric variability as speaker difference, and try to deal with it by employing speaker related information, such as speaker-dependent duration or transformation. However, dysarthric variability is caused by diverse conditions and may be uncertain even for the same speaker. An intuitive idea to deal with dysarthric speech is to explicitly encode the variability and use the encodings as auxiliary features for acoustic modeling. This idea is similar to the speaker aware training \cite{tan2016speaker}. However, unlike the speaker information that can be clearly identified, dysarthric variability is commonly unknown or uncertain, and is difficult to be encoded directly.
In contrast to explicitly model the variability, variational auto-encoder (VAE) \cite{kingma2014stochastic} obtains robust representation by projecting the input features to a low-dimensional latent space, and by applying stochastic gradient variational Bayes (SGVB) algorithm to model uncertainty. Hence variability is suppressed for feature reconstruction. Thanks to this advantage, VAE has been widely used for robust speech recognition \cite{tan2016learning,hsu2017unsupervised,hsu2018extracting}.


In this work, we propose a variational auto-encoder based variability encoder (VAEVE) to explicitly encode the variability for dysarthric speech. The VAEVE makes use of both phoneme information and low-dimensional latent variable to reconstruct the input acoustic features. In this way, the latent variable is forced to encode the phoneme-independent variability instead of suppressing it.
LSTMs are employed in the VAEVE to capture temporal information of dysarthric speech.
SGVB algorithm is applied to model the distribution for generating variability encodings.
For acoustic modeling, encoder of VAEVE associated only with acoustic features is used to generate variability encodings, which are further used as auxiliary input features of DNN acoustic model.
In the ASR task on UASpeech corpus, hybrid DNN systems using the variability encodings consistently outperform the baseline systems, and obtain absolute word error rate (WER) reduction by up to 2.2\% on dysarthric speech with ``Very low'' intelligibility level. The variability encodings are shown to be complementary to speaker information modeled by learning hidden unit contributions (LHUC) \cite{Swietojanski2016} based speaker adaptive training (SAT),
and be effective to the ``Mixed'' type of dysarthric speech with diverse or uncertain conditions.

The rest of the paper is organized as follows. The proposed VAEVE structure and SGVB algorithm is described in Section \ref{sec:VAEVE}.
Section \ref{sec:baseline} presents the baseline DNN system adaptively trained with multiple speakers.
Section \ref{sec:exps} shows the experiment on UASpeech task. The last section draws the conclusion and the future works.

\section{VAE based variability encoding}
\label{sec:VAEVE}

\subsection{Model design}
\label{sec:design}

\begin{figure}[htbp]
  \centering
  \includegraphics[width=7cm,height=3cm]{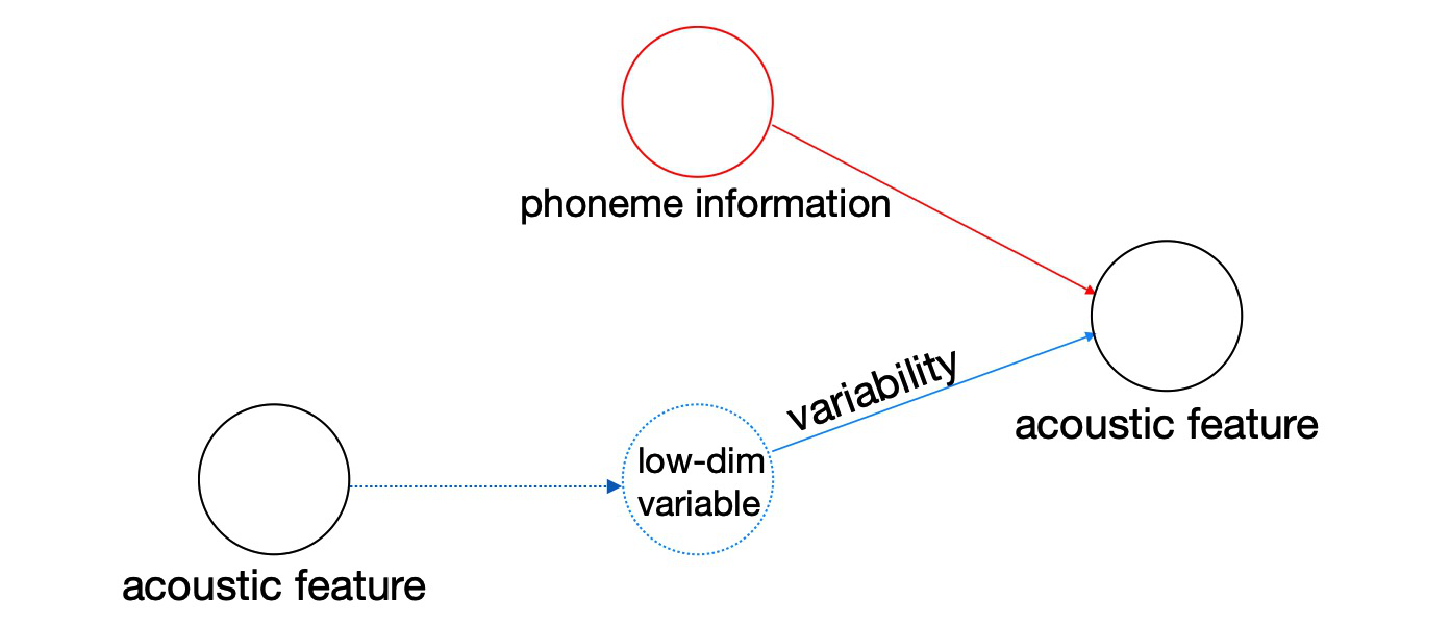}
  \caption{When using both phoneme information and low-dimensional latent variable associated with acoustic feature to generate the acoustic feature, the latent variable would be forced to retain information of acoustic variability.}
  \label{fig:generative}
\end{figure}

Acoustic feature is assumed to manifest the joint effect of phoneme information and acoustic variability. Letting ${\bm o} = \{{\bm o}_{t=1:T}\}$ be acoustic feature sequence, ${\bm c} = \{{\bm c}_{t=1:T}\}$ be phoneme sequence, and ${\bm z} = \{{\bm z}_{t=1:T}\}$ be acoustic variability, the generative model is presented as ${\bm o} \sim p({\bm o}|{\bm c},{\bm z})$.
Unlike speaker information that can be clearly identified, variability of dysarthric speech caused by diverse conditions may be unknown or uncertain. Even modeling certain aspects of the variability, e.g., voice, motion, and tempo, it is still difficult to cover the mixed effect in the acoustic feature.
To encode the acoustic variability explicitly, ${\bm z}$ can be regarded as latent variable generated by an encoder over the acoustic feature. When using a low-dimensional latent space, limited information is encoded from the acoustic feature. Meanwhile, the encoded information is jointly used with the phoneme information to generate the acoustic feature. This would force the encoding to retain more information of phoneme-independent variability. The encoder is similar to that in variational auto-encoder (VAE) \cite{kingma2014stochastic}. This generating process is shown in Figure \ref{fig:generative}, and it leads to the proposed design of variational auto-encoder based variability encoder (VAEVE).

The posterior distribution of ${\bm z}$ can be approximated by the probabilistic encoder output distribution $q({\bm z}|{\bm o},{\bm \phi})$ using SGVB algorithm. The conditional likelihood of ${\bm o}$ is computed as
\begin{eqnarray}
&& \!\!\!\!\!\!\!\!\!\!\!\!\!\!\!\!\!\!\!\! \log p({\bm o}|{\bm c},{\bm \theta}) ~~ = ~~ \log \int p({\bm o},{\bm z}|{\bm c},{\bm \theta}) d{\bm z} ~~ \geq \nonumber \\
&& \!\!\!\!\!\!\!\!\!\!\!\!\!\!\!\!\!\!\!\! \int \!\! q({\bm z}|{\bm o},\!{\bm \phi}) \log p({\bm o}|{\bm c},\!{\bm z},\!{\bm \theta}) d{\bm z} \! - \! KL(q||p) \!\overset{\text{def}}{=}\! \mathcal{L}({\bm \theta},\!{\bm \phi};{\bm o},\!{\bm c})
\label{eq:variational}
\end{eqnarray}
where ${\bm \theta}$ and ${\bm \phi}$ denotes the model parameters of decoder and encoder respectively, and $KL(q||p)$ is the KL divergence between $q({\bm z}|{\bm o},{\bm \phi})$ and prior distribution $p({\bm z})$.
By fixing ${\bm \theta}$, maximizing the lower bound $\mathcal{L}({\bm \theta},{\bm \phi};{\bm o},{\bm c})$ is equivalent to minimize the KL divergence between true posterior $p({\bm z}|{\bm o},{\bm c})$ and $q({\bm z}|{\bm o},{\bm \phi})$. Hence $q({\bm z}|{\bm o},{\bm \phi})$ is an approximation to the true posterior.

\begin{figure}[htbp]
  \centering
  \includegraphics[width=7.5cm,height=3.3cm]{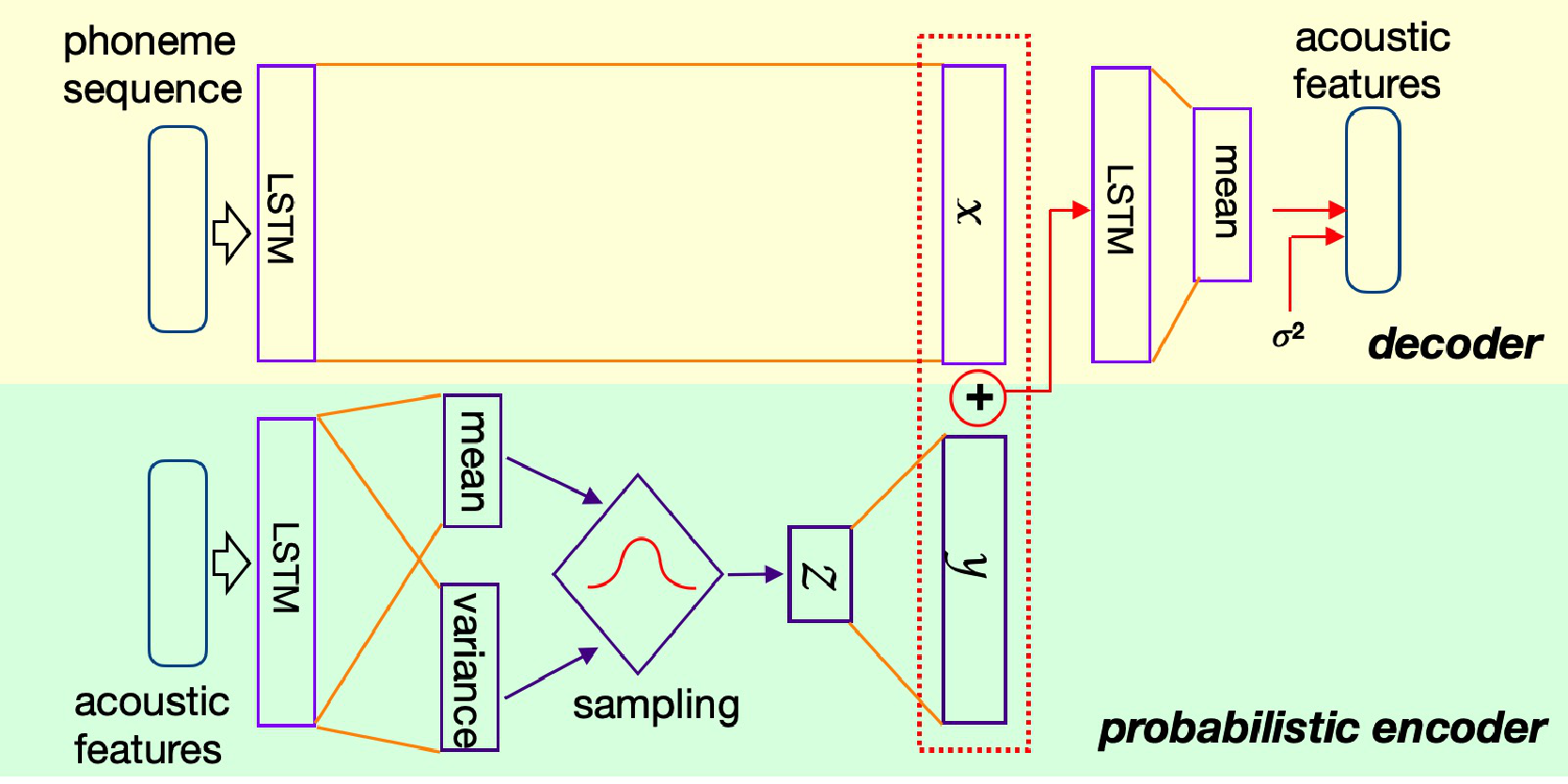}
  \caption{The structure of variational auto-encoder based variability encoder (VAEVE). The upper part is the decoder; the lower part is the encoder.}
  \label{fig:model}
\end{figure}

The probabilistic encoder output distribution $q({\bm z}|{\bm o},{\bm \phi})$ can be computed as a Gaussian distribution with diagonal covariance. For the $t$th instant in an utterance, we have $q({\bm z}_t|{\bm o},{\bm \phi}) = \mathcal{N}({\bm z}_t; {\bm \mu}^{(\text{enc})}_t, ({\bm \sigma}^{(\text{enc})}_t)^2)$. The mean vector and diagonal vector of standard deviation are implemented by an LSTM using the whole utterance as input, such that temporal information can be considered. This is presented as
\begin{eqnarray}
{\bm \mu}^{(\text{enc})}_t & = & {\bm W}^{(\text{enc})}_{\mu} {\bm h}^{(\text{enc})}_t + {\bm b}^{(\text{enc})}_{\mu} \nonumber \\
{\bm \sigma}^{(\text{enc})}_t & = & \exp\{{\bm W}^{(\text{enc})}_{\sigma} {\bm h}^{(\text{enc})}_t + {\bm b}^{(\text{enc})}_{\sigma}\} \nonumber \\
{\bm h}^{(\text{enc})}_t & = & \text{LSTM}^{(\text{enc})}_t \left\{ {\bm o}_{t=1:T} \right\}.
\label{eq:encoder}
\end{eqnarray}
Then, the latent variable ${\bm z}_t$ is sampled as ${\bm z}_t = {\bm \mu}^{(\text{enc})}_t + {\bm \sigma}^{(\text{enc})}_t \otimes {\bm \epsilon}$, where ${\bm \epsilon} \sim \mathcal{N}(0,I)$. In practice, the dimension of latent variable ${\bm z}_t$ should be significantly lower than that of ${\bm h}^{(\text{enc})}_t$. This encoder structure is shown as the lower part in figure \ref{fig:model}.

In the decoder, $p({\bm o}|{\bm c},{\bm z},{\bm \theta})$ is modeled by an isotropic Gaussian distribution. For the $t$th instant we have $p({\bm o}_t|{\bm c},{\bm z},{\bm \theta}) = \mathcal{N}({\bm o}_t; {\bm \mu}^{(\text{dec})}_t, {\sigma}^2I)$. The mean vector is implemented as
\begin{eqnarray}
{\bm \mu}^{(\text{dec})}_t & = & {\bm W}^{(\text{dec})}_{2} {\bm h}^{(\text{dec})}_t + {\bm b}^{(\text{dec})}_{2} \nonumber \\
{\bm h}^{(\text{dec})}_t & = & \text{LSTM}^{(\text{dec}_1)}_t \left\{ {\bf f}_{t=1:T} \right\} \nonumber \\
{\bf f}_t & = & {\bm x}_t + {\bm y}_t, ~~~~~~{\bm x}_t = \text{LSTM}^{(\text{dec}_2)}_t \left\{ {\bm c}_{t=1:T} \right\} \nonumber \\
{\bm y}_t & = & {\bm W}^{(\text{dec})}_{1} \text{Sigmoid}({\bm z}_t) + {\bm b}^{(\text{dec})}_{1} \label{eq:decoder}
\end{eqnarray}
where ${\bm c}_t$ is a one-hot vector representing the $t$th phoneme label. The variance ${\sigma}^2$ is assumed to be a tunable scalar constant. This decoder structure is shown as the upper part in figure \ref{fig:model}.

During training of VAEVE, the prior $p({\bm z})$ is assumed to be $\mathcal{N}(0,I)$. The $\mathcal{L}({\bm \theta},{\bm \phi};{\bm o},{\bm c})$ in equation (\ref{eq:variational}) can be rewritten as
\begin{eqnarray}
&& \!\!\!\!\!\!\!\!\!\!\!\!\! \mathcal{L}({\bm \theta},{\bm \phi};{\bm o},{\bm c}) \! \approx \! \frac{1}{{\sigma}^2} \! \sum_{t=1}^T \! \left\{ \! \frac{{\sigma}^2}{2} \! \sum_{d=1}^D \left( 1 \! + \! 2\log (\!{\sigma}^{(\text{enc})}_{t,d}\!)  -  (\!{\mu}^{(\text{enc})}_{t,d}\!)^{\!2} \! \right.\right. \nonumber \\
&& \!\!\!\!\!\!\!\!\!\!\!\!\! \left.\left. - (\!{\sigma}^{(\text{enc})}_{t,d}\!)^{\!2}\right) - \frac{1}{J} \sum_{j=1}^J \sum_{d=1}^D \frac{1}{2} ({o}_{t,d} - {\mu}^{(\text{dec})}_{t,d,j})^2 + \text{constant} \right\}
\label{eq:loss}
\end{eqnarray}
where $D$ denotes the dimension of latent variable, and $J$ is the number of Monte Carlo sampling. The ${\mu}^{(\text{dec})}_{t,d,j}$ is the $d$th mean element of decoder computed with the $j$th sample of ${\bm z}_t$. During training, the foremost $\frac{1}{{\sigma}^2}$ can be absorbed by learning rate. The lower bound $\mathcal{L}({\bm \theta},{\bm \phi};{\bm o},{\bm c})$ in equation (\ref{eq:loss}) is maximized. The decoder is first pre-trained layer-wise by setting ${\bm y}_t$ to zero. Then, all parameters of VAEVE are fine-tuned jointly.

\subsection{Preventing encoding phoneme information}
\label{sec:operation}

Although low-dimensional latent space is used by probabilistic encoder, it may also try to model frame-level phoneme information greedily such that the VAEVE degenerates into an identity transform of acoustic features. Some operations can be applied to prevent the encoder from encoding frame-level phoneme information.

{\noindent \bf Average Pooling}: The acoustic variability is assumed to be stable in a short period. Average pooling over time axis can force the encoder at each time instant to encode the stable information in a short period. During VAEVE training the average pooling is applied to the ${\bm h}^{(\text{enc})}_t$ in equation (\ref{eq:encoder}), which is rewritten as
\begin{eqnarray}
{\bm h}^{(\text{enc})}_t = \frac{1}{2\tau+1} \sum_{t'=t-\tau:t+\tau} \text{LSTM}^{(\text{enc})}_{t'} \left\{ {\bm o}_{t=1:T} \right\}
\label{eq:average}
\end{eqnarray}
where $\tau$ is the radius of time period.

{\noindent \bf Time Delaying}: Time delaying force the decoder at the $t$th instant to used the latent variable at a preceding instant $t-\Delta t$ as input, where $\Delta t$ is the delayed time. The latent variable ${\bm z}_{t-\Delta t}$ is expected to encode similar information to ${\bm z}_{t}$. The ${\bm y}_t$ of decoder in equation (\ref{eq:decoder}) is rewritten as
\begin{eqnarray}
{\bm y}_t & = & {\bm W}^{(\text{dec})}_{1} \text{Sigmoid}({\bm z}_{t-\Delta t}) + {\bm b}^{(\text{dec})}_{1}.
\label{eq:delay}
\end{eqnarray}

{\noindent \bf Fixing Decoder}: After pre-training, the parameters in the decoder can be fixed. Then, only the encoder is trained in the fine-tuning step. This force the latent variable to encode complementary information to the phoneme sequence.

{\noindent \bf Contextual Input}: The encoder can use successive frames of acoustic features as input instead of one frame at each instant.

\subsection{Variability encoding for DNN acoustic modeling}
\label{sec:acoustic}
%
For acoustic modeling, latent variables generated from the probabilistic encoder of VAEVE is used as variability encodings.
Only the acoustic features are required by the encoder, and no additional information is used when making use of the VAEVE for ASR. During acoustic model training, the means and variances computed by the encoder are used to sample latent variables ${\bm z}$ for each update of DNN parameters. The sampled variables are concatenated with contextual acoustic input features and used as auxiliary input features to train the DNN model. At the $t$th time instant, this is presented as
\begin{equation}
{\bm o}'_t = \text{Concatenation} \{ {\bm o}_{t-\tau}, ..., {\bm o}_{t}, ..., {\bm o}_{t+\tau}, {\bm z}_{t,j} \}
\label{eq:training}
\end{equation}
where $\tau$ is the radius of contextual window, and ${\bm z}_{t,j}$ denotes the sampled latent variable at the $t$th instant for the $j$th update.
However, the latent variables may capture some variability information unrelated to ASR target, such that using ${\bm o}'_t$ to train DNN model from scratch is prone to over-fitting. An approach to addressing the issue is to use ${\bm o}'_t$ to retrain a well-trained DNN model by a few steps. In this strategy, the well-trained model parameters are updated by a small learning rate, and the new weights associated with ${\bm z}_{t,j}$ is updated by a larger learning rate.
In the test stage, the latent variable ${\bm z}_{t,j}$ in equation (\ref{eq:training}) is replaced with the mean vector ${\bm \mu}^{(\text{enc})}_t$ and used for decoding.




\section{Baseline ASR system description}
\label{sec:baseline}

The baseline hybrid DNN acoustic model in this work is the same as that in \cite{genginvestigation,liu2020exploiting}. It consists of seven hidden layers and one output layer. Each hidden layer contains a basic set of neural operations performed in sequence, i.e., affine transformation, ReLU activation, and batch normalization. To reduce the number of model parameters, linear bottleneck projections are applied prior to the second to sixth layers. The outputs of first six layers utilize dropout operation to reduce over-fitting. To accelerate the training process and circumvent the vanishing gradient problem, two skip connections are used to connect the output of the first layer to the third layer, and the output of the fourth layer to the sixth layer respectively.


Multi-task learning (MTL)\cite{Caruana1997} is used to train the DNN system. The labels for the two tasks are based on frame-level tied tri-phone state alignments and mono-phone alignments respectively. Incorporating frame-level mono-phone alignments in the labels reduces the risk of over-fitting to unreliable alignments of frame-level tri-phone states computed from dysarthric speech.

To model the large variability among dysarthric speakers, LHUC based SAT \cite{Swietojanski2016} is employed by the baseline system. Speaker-level LHUC scaling vectors are deployed to the ReLU activation output in the first layer. During training the LHUC vectors are updated once per mini-batch together with the network parameters. Unsupervised LHUC adaptation is performed in the test stage, where the LHUC vectors are updated once per utterance in one epoch.



\section{Experiments and results}
\label{sec:exps}

\subsection{Task description}

The UASpeech \cite{Kim2008a} is an isolated word recognition task consisting of dysarthric speech from 16 dysarthric speakers and normal speech from 13 healthy speakers. The content of speech covers 155 common words and 300 uncommon words. The speakers with dysarthria in types of ``Spastic'', ``Athetoid'', and ``Mixed'' are grouped by their intelligibility levels, which are groups of ``Very low'', ``Low'', ``Mild'', and ``High''. Silence stripping is performed using a GMM-HMM system to remove redundant silence in the recordings as in \cite{Yu2018}. About 30.6 hours of speech from dysarthric and healthy speakers are utilized as training data. About 9 hours of dysarthric speech are used as test data, where 99 words is unseen in training data.

\subsection{System settings}

The hybrid DNN acoustic model is implemented using the Kaldi toolkit \cite{povey2011kaldi}. The contextual input to the DNN acoustic model is 9 successive frames of 80-dimensional filter-bank features with the first order differences. The first six hidden layers contain 2000 neurons each, while the dimension of the linear bottleneck projections is 200 and the dropout rate is 20\%. The seventh hidden layer contains 100 neurons. For multi-task learning, the same weight 0.5 is used for both tasks using the 2001 tied tri-phone states and 41 mono-phones. Cross entropies between the task labels and the DNN outputs are minimized by back-propagation based on RMSProp optimizer. A uniform language model is used for decoding. Significance test are based on the matched pairs sentence-segment word error approach.

The VAEVE is also implemented based on the Kaldi toolkit. For the probabilistic encoder in VAEVE, the LSTM contains 128 cells, and the dimension $D$ of latent variable ${\bm z}_t$ is 39. The acoustic features as encoder input are the same filter-bank features used as DNN acoustic model input. Each LSTMs in the decoder has 256 cells. The mono-phone alignments serve as the phoneme sequences in the decoder. In equation (\ref{eq:loss}) for VAEVE training, we empirically set ${\sigma} = 0.01$ and $J=1$. RMSProp optimizer based back-propagation is used for VAEVE training.

When using the generated latent variable ${\bm z}_t$ for acoustic modeling, the new input features ${\bm o}'_t$ in equation (\ref{eq:training}) is used to retrain the well-trained DNN acoustic model with fixed batch normalization. The retraining learning rate of the well-trained parameters is initialized by the final learning rate in DNN training. The retraining learning rate of the new weights associated with ${\bm z}_t$ is initialized as 100 times of that of well-trained parameters. Four epochs are performed with the initialized learning rate, which is then halved in the following epochs.

\subsection{Results on speaker-independent DNN system}


Systems (1) to (3) in Table \ref{tab:intel} show the performance of speaker-independent DNN systems with or without using variability encodings generated from VAEVE. In the VAEVE using average pooling, $\tau$ in equation (\ref{eq:average}) is set to 10 frames. Consistent performance improvement is obtained by the DNN systems using variability encodings (Sys (2) and (3)) compared to the baseline system (Sys (1)). Although the VAEVEs with or without using average pooling achieve the same overall WER, they yield different improvements in different intelligibility levels. It seems that the VAEVE with average pooling (Sys (3)) performs better in groups with lower intelligibility than the original VAEVE (Sys (2)) without using operation presented in Section \ref{sec:operation}.
\begin{table}[htbp]
\begin{center}
  {\caption{Performance of applying variability encodings generated from VAEVE to the DNN systems on the 16 UASpeech dysarthric speakers. ``Origin'' refers to VAEVE without using operation presented in Section \ref{sec:operation}, and ``Average'', ``Delay'', ``FixDec'', and ``Cntxt'' refer to VAEVEs applying average pooling, time delaying, fixing decoder, and contextual input. ``Very low'', ``Low'', ``Mild'' and ``High'' refer to groups with different intelligibility levels. ``\dag'' means the improvement over the comparable baseline system is significant ($P \leq 0.05$). }
    \scalebox{0.71}[0.75]
    {
  \begin{tabular}{c|c|c|c|c|c|c|c|c}
  \hline\hline
   \multirow{2}{*}{Sys} & \multirow{2}{*}{VAEVE} & LHUC & Data & \multicolumn{5}{c}{WER(\%)} \\ \cline{5-9}
                        &                        & SAT & Aug. & Very low & Low & Mild & High & Overall \\ \hline \hline
   (1)                  &  -                     & $\times$ & $\times$ & 69.8 & 32.6 & 24.5 & 10.4 & 31.5 \\ \hline
   (2)                  & Origin                 & $\times$ & $\times$ & 69.0\dag & 32.6 & {\bf 23.6}\dag & {\bf 9.8}\dag & {\bf 30.9}\dag \\
   (3)                  & Average                & $\times$ & $\times$ & {\bf 68.6}\dag & {\bf 32.5} & 23.9 & 10.0\dag & {\bf 30.9}\dag \\ \hline \hline
   (4)                  &  -                     & $\surd$ & $\times$ & 64.4 & 29.9 & 20.3 & 9.0 & 28.3 \\ \hline
   (5)                  & Origin                 & $\surd$ & $\times$ & 63.3\dag & 29.1\dag & 19.3\dag & 8.6\dag & 27.5\dag \\
   (6)                  & Average                & $\surd$ & $\times$ & {\bf 62.2}\dag & 28.8\dag & 19.8 & 8.6\dag & 27.3\dag \\
   (7)                  & Delay                  & $\surd$ & $\times$ & 62.9\dag & 29.0\dag & 19.2\dag & 8.6\dag & 27.4\dag \\ \cline{2-9}
   (8)                  & FixDec                & $\surd$ & $\times$ & 62.9\dag & 28.7\dag & 19.0\dag & 8.6\dag & 27.3\dag \\
   (9)                  &  + Delay        & $\surd$ & $\times$ & 62.6\dag & {\bf 28.6}\dag & 19.0\dag & 8.7\dag & {\bf 27.2}\dag \\
   (10)                  &  + Cntxt        & $\surd$ & $\times$ & 62.8\dag & 28.7\dag & {\bf 18.8}\dag & {\bf 8.5}\dag & {\bf 27.2}\dag \\ \hline \hline
   (11)                  &  -                     & $\surd$ & $\surd$ & 62.4 & 27.6 & 17.4 & {\bf 7.9} & 26.4 \\ \hline
   (12)                  & Average                & $\surd$ & $\surd$ & {\bf 61.2}\dag & {\bf 26.3}\dag & {\bf 16.8}\dag & 8.0 & {\bf 25.7}\dag \\ \hline\hline
  \end{tabular}
  }
  \label{tab:intel}
  }
\end{center}
\end{table}


\subsection{Results on DNN system trained with LHUC SAT}

Performance of applying variability encodings generated from VAEVE to the DNN systems trained with LHUC SAT is presented as Systems (5) to (10) in Table \ref{tab:intel}. For the VAEVE using time delaying, $\Delta t$ in equation (\ref{eq:delay}) is set to 10 frames. For that using contextual input, 9 successive frames of filter-bank features are used as encoder input. Consistent performance improvement is achieved by the DNN systems using variability encodings (Sys (5) to (10)) compared to the baseline system (Sys (4)). The overall improvement is up to 1.1\% (Sys (10)) in absolute WER reduction, which is more significant than that for speaker-independent DNN. This implies that the variability encodings generated by VAEVE are complementary to the speaker information modeled by LHUC SAT. Among these systems, the VAEVE applying average pooling (Sys (6)) performs the best on the group with ``Very low'' intelligibility. The absolute WER reduction is 2.2\% over the baseline system. The joint use of fixing decoder and contextual input for VAEVE (Sys (10)) obtains better performance on groups with higher intelligibility, and achieves the best overall performance.

Table \ref{tab:type} shows the performance of using VAEVE based variability encodings on speakers with different dysarthria types. It shows that on the ``Mixed'' type of dysarthric speech with diverse or uncertain conditions, using the variability encodings gives a statistically significant WER reduction by up to 2.0\% (Sys (6)) over the baseline system (Sys (4)). This suggests the effectiveness of the VAEVE method for variability modeling.
\begin{table}[htbp]
\begin{center}
  {\caption{Performance of applying VAEVE based variability encodings on the UASpeech speakers with different dysarthria types including ``Spastic'', ``Athetoid'', and ``Mixed''. }
    \scalebox{0.8}[0.75]
    {
  \begin{tabular}{c|c|c|c|c}
  \hline\hline
   \multirow{2}{*}{System in Table \ref{tab:intel}} & \multirow{2}{*}{VAEVE} & \multicolumn{3}{c}{WER(\%)} \\ \cline{3-5}
                         &  & Spastic & Athetoid & Mixed \\ \hline\hline
   Sys (4) & -                     & 28.8 & 21.5 & 32.3 \\\hline
   Sys (6) & Average                 & 27.9\dag & 21.1 & {\bf 30.3}\dag \\
   Sys (10) & FixDec+Cntxt                 & {\bf 27.8}\dag & {\bf 20.0}\dag & 31.0\dag \\ \hline\hline
  \end{tabular}
  }
  \label{tab:type}
  }
\end{center}
\end{table}

In addition, the variability encoding is applied to the system trained with data augmentation, which is shown as System (12) in Table \ref{tab:intel}. The data amount is augmented to 207.5 hours by speed perturbation introduced in \cite{genginvestigation}.
To apply variability encoding, latent variables are sampled for the training data without augmentation. Then, the concatenated features in equation (\ref{eq:training}) of the training data is employed to retrain the baseline system trained with data augmentation. No augmented data is really used in the retraining applying variability encoding. This enables the retraining process to benefit from data augmentation while without significantly enlarging the retraining time. LHUC adaptation is then applied to test data. Consistent performance improvement is obtained by the system applying variability encoding (Sys (12)) compared to the baseline system (Sys (11)).
Table \ref{tab:all} presents the performance of published systems and our system shown as System (12) in Table \ref{tab:intel} on UASpeech task.
\begin{table}[htbp]
\begin{center}
  {\caption{A comparison on UASpeech task between published systems and our system (Sys (12) in Table \ref{tab:intel}). }
    \scalebox{0.80}[0.75]
    {
  \begin{tabular}{c c}
  \hline
   Systems & WER (\%) \\ \hline
   Sheffield-2013 Cross domain augmentation \cite{christensen2013combining} & 37.5 \\
   Sheffield-2015 Speaker adaptive training \cite{sehgal2015model} & 34.8 \\
   CUHK-2018 DNN System combination \cite{Yu2018} & 30.6 \\
   Sheffield-2019 Kaldi TDNN + Data Aug. \cite{xiong2019phonetic} & 27.9 \\
   CNN-TDNN + Speaker adaptation + Transfer learning \cite{xiong2020source} & 30.8 \\
   DNN + Data Aug. + LHUC SAT + Cross domain visual \cite{liu2020exploiting} & 26.8 \\
   DNN + Data Aug. + LHUC SAT \cite{genginvestigation} & 26.4 \\
   LAS + CTC + Meta-learning + SAT \cite{wang2021improved} & 35.0 \\
   QuartzNet + CTC + Meta-learning + SAT \cite{wang2021improved} & 30.5 \\
   {\bf DNN + Data Aug. + LHUC SAT + AVEVE (ours)} & {\bf 25.7} \\ \hline
  \end{tabular}
  }
  \label{tab:all}
  }
\end{center}
\end{table}

\section{Conclusion}
\label{sec:conclus}
%
%

This work introduces a variational auto-encoder based variability encoder (VAEVE) to encode the variability for dysarthric speech. Both phoneme information and low-dimensional latent variable are used to reconstruct the input acoustic features in VAEVE, such that the latent variable is forced to encode variability information. SGVB algorithm is used for learning VAEVE.
For acoustic modeling, the variability encodings are generated and used as auxiliary features.
In the speech recognition task on UASpeech corpus,
up to 2.2\% absolute WER reduction on dysarthric speech with ``Very low'' intelligibility is obtained by using variability encodings.
The future works may focus on applying VAEVE to acoustic modeling of noisy data.

\section{Acknowledgements}

This work is supported by Natural Science Foundation of China U1736202, and Shenzhen Fundamental Research Program JCYJ20160429184226930 and KQJSCX20170731163308665.


%
%


%

\newpage
\bibliographystyle{IEEEtran}
\bibliography{refs_xxr_checked}

\end{document}